\documentclass[onecolumn, revtex-4, tighten]{openjournal}

\usepackage{graphicx}
\usepackage[utf8]{inputenc} 
\usepackage[T1]{fontenc}    
\usepackage[hidelinks]{hyperref}       
\usepackage{url}            
\usepackage{booktabs}       
\usepackage{amsfonts}       
\usepackage{nicefrac}       
\usepackage{microtype}      
\usepackage{amsmath, amssymb}

\begin{document}

\title{Hawking Radiation from non-evaporating primordial black holes cannot enable the formation of direct collapse black holes}

\author{Jonathan Regan$^{1,2,*}$ \href{https://orcid.org/0009-0001-6521-5884}{\includegraphics[scale=0.6]{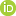}}} 
\author{Marios Kalomenopoulos$^{2}$ \href{https://orcid.org/0000-0001-6677-949X}{\includegraphics[scale=0.6]{Images/ORCIDiD_icon.png}}}
\author{Kelly Kosmo O'Neil$^{1}$ \href{https://orcid.org/0000-0003-2400-7322}{\includegraphics[scale=0.6]{Images/ORCIDiD_icon.png}}}

\thanks{$^*$E-mail: jonathan.regan@unlv.edu, jonathanregan12@gmail.com}

\affiliation{$^{1}$ Department of Physics, University of Nevada, Reno\\ 
  $^{2}$ Department of Physics and Astronomy, University of Nevada, Las Vegas}

\begin{abstract}

The formation of supermassive black holes (SMBHs) in the early Universe is a subject of significant debate. In this study, we examine whether non-evaporating primordial black holes (PBHs) can offer a solution. We establish initial constraints on the range of PBH masses that correspond to Hawking radiation (HR) effective temperatures in the range needed to suppress $H_2$ cooling, which would facilitate the formation of massive black hole seeds in atomic cooling halos. We also investigate the specific intensity of the HR from non-evaporating PBHs and compare it with the critical radiation needed for direct collapse black holes (DCBHs). We show that HR from non-evaporating PBHs cannot serve as an irradiating mechanism to facilitate the formation of the seeds for the SMBHs we observe in the high-redshift Universe unless, perhaps, the PBHs within the relevant mass range comprise a significant fraction of dark matter and are significantly clustered towards the center of the primordial halo.

\end{abstract}


\maketitle


\section{Introduction}\label{sec:Introduction}

In the standard model of cosmology, our Universe is expanding from a hot and dense state \citep[e.g.][]{PDD_LCDM}. Astrophysical structures begin to form when initial fluctuations collapse to form the first gas clouds (also known as primordial gas clouds). These are the birthplaces of the first galaxies \citep[e.g.][]{Johnson_Claudio_Khochfar_2012, Yajima_et_al_2017, Cimatti_2020, Volonteri_et_al_2021}. Many, if not all, of these galaxies are believed to harbor a massive black hole (BH) in their center. Observations of quasars formed in the early Universe indicate that a number of these BHs have masses of more than a billion solar masses, and were formed before the Universe was a billion years old\footnote{Such as the quasar ULAS J1342+0928, whose SMBH has a mass of 800 million solar masses \citep{Banados_2017}. See also \citep{Wu_et_al_2015, Mortlock_et_al_2011, Fan_et_al_2003, Inayoshi_2020}.}. The formation of these high-mass black holes, aptly called supermassive black holes (SMBHs), is a matter of considerable debate \cite[e.g.][]{Rees_1984, Inayoshi_2020}. 

Three of the most prevalent proposed scenarios, outlined in \cite{Rees_1984}, include: 1) the collapse of massive Pop III stars and their growth due to accretion \citep{Omukai_2001, Yajima_Khochfar_2016,  Klessen_2019, Latif_Whalen_Khochfar_2022}, 2) runaway mergers in dense star clusters \citep{Katz_2019}, and 3) the `direct collapse' scenario. There are many mechanisms proposed for the latter, including the abrupt collapse of a metal-free (or metal-poor) gas cloud due to the suppressing role of Lyman-Werner (LW) radiation on molecular cooling due to $H_2$, dynamical heating, streaming motions of baryon-dark matter, the increase of the mass of protostellar collapse in the presence of magnetic fields, or gas dynamical processes \citep{Begelman_et_al_2006, Lodato_Natarajan_2006, Schleicher_et_al_2009, Begelman_2010, Shang_Bryan_Haiman_2010, Agarwal_et_al_2012, Agarwal_et_al_2014_billion, Hirano_et_al_2017, Agarwal_2019, Latif_2019_inflow, Latif_et_al_2021, Hirano_Machida_Basu_2021, Latif_et_al_2022, Latif_Schleicher_Khochfar_2023, Regan_Volonteri_2024, Chon_Omukai_2024}. Alternative scenarios, among others, involve the growth of primordial black holes (PBHs) \citep{Bean_Magueijo_2002, Duchting_2004}. For a more detailed coverage of these and other scenarios, see \citep{Volonteri_2010, Latif_Ferrara_2016, Becerra_2018, Smith_Bromm_2019, Inayoshi_2020, Haemmerle_et_al_2020, Volonteri_et_al_2021}.

Even the most common proposals are not without problems. For example, a newly-born Pop III star in the range of  $\sim\rm 100 \: M_\odot$ would need to accrete continuously at a rate close to the Eddington limit for at least a Gyr to reach masses close to $\rm 10^9 \: M_\odot$ at redshifts $z= 6-7$ \citep{Milosavljevic_et_al_2009, Latif_Ferrara_2016, Smith_Bromm_2019}, which poses difficulties for the first scenario. Dense star clusters would need to be able to retain their members after the mergers (i.e. to overpower the gravitational kicks that lead to the ejection of the merged object from the cluster \citep{Merritt_et_al_2004, Loeb_Furlanetto_2013}), posing difficulties for the second scenario.

This paper focuses on the third scenario: the direct collapse of metal-poor gas clouds, a `heavy seeds' scenario. In this case, the masses of the initial BHs can reach up to $\rm 10^5-10^6 \: M_\odot$ \cite[e.g.][]{Agarwal_et_al_2014_billion, Woods_et_al_2017} due to general relativistic instabilities\footnote{We note though that recent simulation results \citep{Regan_Volonteri_2024, Liu_et_al_2024} show that there is not a clear-cut, bimodal distribution between `heavy' and `light' seeds, and there are mechanisms that produce heavy BHs as the tails of a continuum mass spectrum.}. Therefore, they could grow through more viable rates of accretion into the SMBHs we observe by $z \approx 6-7$. One of the usual prerequisites for this scenario is to avoid fragmentation of the initial gas cloud to many smaller stars, and instead form directly a single, supermassive star that would lead to a massive BH\footnote{The presence of angular momentum is also an obstacle to massive BH collapse that needs to be overcome. Here, we assume that the gas clouds in question have overcome this issue by assuming they have either formed with negligible angular momentum, or that they have lost their initial angular momentum due to interactions with their environment \citep{Koushiappas_et_al_2004} and/or primordial magnetic seed fields \citep{Hirano_Machida_Basu_2021}}. In reality, the situation is more complex, and it is still uncertain whether fragmentation can be completely avoided or if it is a basic prerequisite for heavy seed formation \citep{Klessen_Glover_2023, Liu_et_al_2024}. Studies have shown that it is still possible to form a massive BH seed, even in the presence of moderate fragmentation. The key in these cases is high accretion rates (i.e. $\dot{M} > 0.1 M_\odot/{\rm year}$) of the protostars so that they can grow rapidly with negligible radiative feedback and an enhanced cross-section for collisions, which would allow the merging of numerous fragments \citep{Omukai_Palla_2001, Reinoso_et_al_2023, Boekholt_et_al_2018} (for some alternative direct-collapse scenarios see also e.g. \citep{Mayer_Bonoli_2019, Latif_et_al_2022}). In all cases, it is highly preferential that the collapse of the gas be delayed to happen in an `atomic cooling cloud` - a cloud without significant metals or molecules that could cool efficiently - so that high accretion rates can be achieved. Such a cloud could form if production of $H_2$, the main coolant in the early Universe, is suppressed. This can be accomplished by a number of mechanisms \citep{Inayoshi_2020, Regan_et_al_2020, Wise_et_al_2019, Visbal_et_al_2014}. In this paper, we consider the presence of strong UV radiation in the LW range ($11.2-13.6$ eV) that is able to suppress the formation of molecular hydrogen H$_2$ \citep{Omukai_2001, Oh_Haiman_2002, Bromm_Loeb_2003, Agarwal_Khochfar_2015, Agarwal_et_al_2019_cooling} through radiation from PBHs.

For this mechanism, we investigate whether the Hawking radiation (HR) \citep{Hawking_1974} emitted from early-Universe PBHs \citep{Hawking_1971_PBHs, Carr_Hawking_1974_PBHs} can create the necessary conditions for the formation of large BH seeds via the direct collapse of primordial gas clouds. In this proposed mechanism, the required photons are supplied by the HR of an evaporating PBH that is still far from ``exploding''. We model this radiation as a blackbody spectrum, focusing only on primary photon emission.

The role of PBHs in the formation of the first stars and galaxies has been investigated before \citep{Liu_Zhang_Bromm_2022, Liu_Bromm_2023, Lu_Picker_Kusenko_2024}, but remains an important unanswered question in the scientific community. Most recently, \cite{Lu_Picker_Kusenko_2024} studied the heating effects of secondary HR spectra from ``exploding'' PBHs in primordial gas clouds. They find that in the presence of significant clustering of the PBHs in the clouds, PBHs of masses on the order of $10^{14}$ g can lead to a successful direct collapse of the cloud. On the other hand, \cite{Liu_Zhang_Bromm_2022, Liu_Bromm_2023} used cosmological simulations and semi-analytical models to explore the effects of stellar-mass ($\sim 10-100\ \rm M_\odot$) PBHs in the acceleration of structure formation and gas heating by accretion feedback on the BHs.  In the former case, they find that structures are similar to the ones found in $\Lambda$CDM simulations, but in latter case, they show that LW photons from accretion onto stellar-mass PBHs in the halo can indeed promote the formation of massive seeds through the direct-collapse scenario. In this work, we describe a simplified and easily-modifiable analysis framework, more in line with \cite{Lu_Picker_Kusenko_2024}, in hopes that the scientific community can continue to build on this work to answer this open question.

For all cosmological calculations in this paper, we assume a flat $\Lambda$CDM cosmology with $H_0 = 70$ km/s/Mpc and $\Omega_m = 0.3$. This paper is organized as follows: Section \ref{sec:Hawking_radiation} reviews the main consequences of Hawking radiation. Section \ref{sec:Primordial_BHs} introduces some of the relevant features of primordial black holes. Section \ref{sec:PBHs_constraints} sets constraints on the primordial black holes that could act as sources of the Hawking radiation needed to produce `heavy seeds', and compares these constraints with current observational limits. Finally, Section \ref{sec:Discussion} provides a comprehensive discussion of the assumptions and caveats of this work and proposes extensions to this simplified model.


\section{Hawking Radiation}\label{sec:Hawking_radiation}

This section provides a review of the main consequences of HR from a BH. We assume a Schwarzschild BH, ignoring spins which are expected to be very small in most cases for PBHs \citep{DeLuca_et_al_2019, Mirbabayi_Gruzinov_Norena_2020} that we consider in Section \ref{sec:Primordial_BHs}, and consider only its primary photon emission. For a discussion on the validity of these assumptions, see Section \ref{sec:Discussion}.

\subsection{Blackbody spectrum}

The main component of HR can be modeled as a blackbody spectrum,

\begin{equation}\label{eq:Planck_law}
B_{\nu}=f_{\Gamma}f_{\rm ph}\frac{2h\nu^3}{c^2}\frac{1}{\exp[h\nu/k_{B}T]-1}, 
\end{equation}

where all the constants and variables have their usual meaning\footnote{More specifically: $\nu$ is the frequency of emitted photons, $T$ is the temperature of the blackbody, $h$ is Planck's constant, $c$ is the speed of light, and $k_{B}$ is Boltzmann's constant.} and $f_{\rm ph}$ refers to the percentage of HR that is in photons, with $f_{\rm ph}=1$ denoting the case where all of the HR is in photons. We use $f_{\rm ph}=0.2$ \citep{Page_1976}, unless stated otherwise. $f_{\Gamma}$ corresponds to a constant `greybody' factor of $f_{\Gamma}=0.24$, as described in Section \ref{sec:greybody_factors}.

According to Hawking's analysis \citep{Hawking_1974}, the temperature of a BH is connected to its mass via the expression

\begin{equation}\label{eq:Temperature_HR}
T_{\rm BH}=\frac{\hbar c^3}{8\pi k_{B}GM_{\rm BH}} \simeq 6.17 \times 10^{-8} \left( \frac{M_\odot}{M_{\rm BH}}\right)\ {\rm K},
\end{equation}

where $M_{\rm BH}$ is the mass of the black hole and $G$ is Newton's gravitational constant. Hence, the temperature of a BH is inversely proportional to its mass. This is a feature we will exploit in Section \ref{sec:PBHs_constraints}.

Eqs. (\ref{eq:Planck_law}) and (\ref{eq:Temperature_HR}) can then be combined to get the blackbody spectrum as a function of mass:

\begin{equation}\label{eq:Blackbody_spectrum}
B_{\nu}=f_{\Gamma}f_{\rm ph}\frac{2h\nu^3}{c^2}\frac{1}{\exp[16\pi^2 G M_{\rm BH}\nu /c^3]-1} .
\end{equation}

\subsection{Mass evolution and evaporation time}

The energy of the emitted radiation comes from the BH, which in turn ``loses'' its mass. As a result, the temperature of the blackbody increases, as does the peak energy of radiation, leading to increased mass loss. Equating the energy emitted to the Stefan-Boltzmann luminosity provides the evolution of the BH mass with time:

\begin{equation}\label{eq:t_HR_diff_eq}
-\frac{dM_{\rm BH}}{dt}c^2 = L \Rightarrow \frac{dM_{\rm BH}}{dt} c^2=-A \sigma T^4 = -16 \pi \left( \frac{GM_{\rm BH}}{c^2} \right)^2 \left(\frac{\hbar c^3}{8 \pi G M_{\rm BH} k_B}\right)^4,
\end{equation}
where $A$ is the Schwarzschild surface area and $\sigma$ is the Stefan-Boltzmann constant.

Solving eq. (\ref{eq:t_HR_diff_eq}), assuming $t_{\rm initial} \rightarrow 0$, the time evolution of the BH mass since the Big Bang is:

\begin{equation}\label{eq:HR_mass_evolution}
    M_{\rm BH}(t) = \left[M_i^3 - \frac{3}{256} \frac{c^6 \sigma}{\pi^3 G^2} \left(\frac{\hbar}{k_B} \right)^4 t \right]^{1/3} = \left[M_i^3 - C \cdot t(z, H_0, \Omega_m) \right]^{1/3},
\end{equation}
with

\[
C = 4.76 \times 10^{-59}\ M_\odot^3/{\rm Gyr}.
\]

The very small value of this constant indicates that there are no significant changes in mass until the very final moments of BH evaporation. 

The evaporation time $t_{\rm ev}$ of a BH is the amount of time required for it to completely evaporate due to the emission of HR. Setting $M_{\rm BH}(t_{\rm ev})=0$ in eq. \ref{eq:HR_mass_evolution}, it is straightforward to find:

\begin{equation}\label{eq:Evaporation_time}
t_{\rm ev}= \frac{15360 \pi G^2 M_{\rm BH}^3}{3 \hbar c^4} = 2.1\times10^{67}\, \left(\frac{M_{\rm BH}}{M_\odot}\right)^3 \: {\rm yr}.
\end{equation}

As such, the maximum mass of a BH that could have evaporated due to HR within the age of the Universe ($\rm \sim 13.7 \: Gyr$) is $\rm \sim 9 \times 10^{-20} \: M_{\odot}$ (or $\sim 1.8 \times 10^{14}$ g). If one takes into account the larger variety of particles emitted as a BH evaporates (other than photons), then the evaporation time becomes shorter \citep{Mosbech_Picker_2022}. We comment more on this in Section \ref{sec:Discussion}.

Finally, the time evolution of HR temperature is given by inserting the mass evolution, eq. (\ref{eq:HR_mass_evolution}), into the HR temperature, eq. (\ref{eq:Temperature_HR}):

\begin{equation}\label{eq:T_evolution_HR}
    T = \frac{\hbar c^3}{8\pi k_{B}G}\frac{1}{\left[M_i^3 - C \cdot t(z, H_0, \Omega_m) \right]^{1/3}}.
\end{equation}

For the BHs of interest in this work, the temperatures remain effectively constant over relevant timescales, as confirmed in Figure (\ref{fig:mass_evolution_redshift}). As such, we ignore the effects of time evolution in the analysis from this point on.

\section{Primordial Black Holes}\label{sec:Primordial_BHs}

This work investigates whether HR from PBHs can act as an irradiation source of primordial gas clouds, facilitating their collapse into DCBHs. Since the latter demands temperatures of the order of $10^4\: {\rm K}$, this requires BHs with masses much smaller than $\rm 1 \: M_\odot$ (see Section \ref{sec:PBHs_constraints}). A natural candidate for the source of HR in this scenario is primordial black holes.


PBHs are standard black holes formed in the early stages of the Universe when density fluctuations surpass a specific threshold \cite[e.g.][]{Hawking_1971_PBHs, Carr_Hawking_1974_PBHs, Carr_1975_PBHs_spectrum, Carr_Kuhnel_2020}. PBHs are well known as possible dark matter candidates, and most importantly ones that do not require any "new physics" to explain. The birth, lifespan, and ultimate collapse of a PBH are all based on concepts of general relativity. PBHs may also contribute to a number of other phenomena including lensing, gravitational wave events, and growth of structure in the high-redshift Universe \cite[e.g.][]{Escriva_et_al_2022_PBH}. 

The (maximal) mass of a PBH at formation time, assuming a period of a radiation-dominated universe, is comparable to the horizon mass \citep{Byrnes_Cole_2021_PBH, Arbey_2024_PBH}:
\begin{equation}
    M_{\rm PBH} \sim M_H = \frac{4}{3}\pi \rho \left( \frac{1}{H}\right)^3 \sim 10^{15} \left( \frac{t}{10^{-23}{\rm s}} \right) {\rm g},
\end{equation}

where $t$ is the time since the Big Bang. The reference mass of $10^{15}$ g corresponds to a PBH that will be evaporating today due to HR. In contrast, at $t\sim 1$ s (redshift $z \sim 10^9$ based on $\Lambda$CDM), the mass of a PBH can be on the order of $\sim 10^5 \: \rm M_\odot$. As such, the range of possible PBH masses spans many orders of magnitudes.

PBHs have not yet been observed, but a number of constraints have been placed on them through indirect methods either via their gravitational effects or due to their radiation effects from HR and accretion \cite[e.g.][]{Rice_Zhang_2017, Carr_et_al_2021, Green_Kavanagh_2021, Auffinger_2023}. These constraints are generally phrased in terms of the fraction of dark matter (DM) occupied by PBHs, $f_{\rm PBH} = \rho_{\rm PBH}/\rho_{\rm DM}$, where $f_{\rm PBH}=1$ would correspond to the PBHs making up all of the DM. The filled regions of Figure \ref{fig:PBHbounds_DCBHs_Tz} show some of the main observational constraints from \cite{Green_Kavanagh_2021}. Excluding the range of masses where $f_{\rm PBH}$ remains effectively unconstrained ($\sim 10^{17}-10^{21}$ g or $10^{-17}-10^{-12}\ \rm M_\odot$, as well as the most massive PBHs), $f_{\rm PBH} < 0.1$ in all of the other mass ranges. Note that the constraints are model dependent, and the numbers quoted above are optimistic limits (for some ranges, $f_{\rm PBH} \ll 1$).

\section{HR from PBHs as a source of DCBHs}\label{sec:PBHs_constraints}

The key criteria required for the formation of DCBHs in the scenario considered here include:

\begin{itemize}
    \item The presence of radiation that would be able to dissociate or suppress the formation of molecular $H_2$, the main coolant in the early Universe \citep{Agarwal_2019}. This can be achieved with strong UV radiation in the LW range ($11.2-13.6$ eV) directly, and indirectly with the destruction of $H^-$, a dominant intermediate species in the formation of $H_2$, by irradiation of photons with energy $E_\gamma \gtrsim 0.8$ eV \citep{Sugimura_Omukai_Inoue_2014}. We discuss both cases below (Sections \ref{sec:Xray_constraint} and \ref{sec:Tcrit_constraint}), together with the contribution of other radiation feedback. Our main emphasis is in the LW band.

    \item The presence of radiation with a strength of $J_{\rm crit} \sim 10 - 10^5$ in conventional units of $10^{-21}$ erg/s/cm$^2$ /Hz/sr. In practice, we do not use a universal $J_{\rm crit}$, but a critical value that depends on the radiation temperature, following the prescription in \cite{Sugimura_Omukai_Inoue_2014}. For more details, see Section \ref{sec:CRI}.

    \item The above contribute to the delayed collapse of a gas cloud at redshifts in the range [$z\ \epsilon \ (10, 30)$], inside a massive atomic cooling halo, where significant accretion rates are possible. Then, the formation of a massive BH seed is likely.
\end{itemize}

In this section, we investigate whether we can satisfy the above criteria with the presence of PBHs. In other words, in the redshift range of interest [$z\ \epsilon \ (10, 30)$], is it feasible for a population of PBHs to emit blackbody radiation with effective temperature $T_{\rm crit}$ and with enough intensity in the frequency ranges of interest to surpass $J_{\rm crit}$?

\subsection{Constraints from evaporation redshift}\label{sec:Zevap_constraint}

The scenario investigated in this work requires that the PBHs have not evaporated before $z_{\rm min} = 10$. This sets a lower limit to the masses of PBHs of interest, i.e. $M_{\rm PBH} > M_{\rm evap}$. Assuming the fiducial cosmological model, $z=10$ corresponds to an age of the universe $t_{\rm age}(z=10) = t_{\rm age, 10} \sim 0.47$ Gyr. Equating this with evaporation time, and demanding $t_{\rm evap} \gtrsim t_{\rm age, 10}$, we get a mass constraint: $M_{\rm evap} \gtrsim 2.8 \times 10^{-20} \: \rm M_\odot$ (or $5.5 \times 10^{13}$ g).

\subsection{Constraints from X-ray feedback}\label{sec:Xray_constraint}

The effect of X-rays in the chemistry of the gas cloud is complicated, especially when it is present simultaneously with a LW background \citep{Ricotti_2016, Park_Ricotti_Sugimura_2021}. On the one hand, they can provide positive radiative feedback to star formation by enhancing production of $H_2$, and hence cooling. On the other, an intense X-ray background can provide heating that prevents gas collapse \citep{Haiman_Abel_Rees_2000, Park_Ricotti_Sugimura_2021}.

As a first step to estimate these effects, we find the PBH masses where LW photons are stronger than X-ray photons by calculating the ratio $B_{\rm LW}(M_{\rm BH})/B_{\rm X-rays}(M_{\rm BH}) \gtrsim 1000$. $B_{\rm LW}$ denotes the (blackbody) specific intensity at the center of the LW band ($12.5$ eV), and $B_{\rm X-rays}$ denotes the specific intensity of X-ray photons at an energy of $1$ keV and $0.2$ keV. This leads to a mass constraint: $M_{\rm x-ray, 1 keV} \gtrsim 1.0 \times 10^{-13} \: \rm M_\odot$ (or $2.0 \times 10^{20}$ g) and $M_{\rm x-ray, 0.2 keV} \gtrsim 4.0 \times 10^{-13} \: \rm M_\odot$ (or $8.0 \times 10^{20}$ g). This constraint poses a more stringent lower limit than the evaporation constraint described above. 

For our constraints, we used estimates for the ratio of specific intensities and energies (which in our case are equivalent since we consider the PBHs as the sources of both) from Figure $3$ of \cite{Ricotti_2016} and Figure $5$ of \cite{Park_Ricotti_Sugimura_2021}. Both papers study situations of realistic minihalos in a joint LW and X-ray background, and generally find that the presence of strong X-ray radiation can lead to faster fragmentation and production of multiple Pop III stars in lower-mass halos. Note that the ratios they report correspond to a LW background of significant intensity, i.e. $J_{\rm LW, 21} > 10^{-2}$, which is much bigger than what we find below. Nevertheless, in our case we use their estimates simply as a guide to the masses of interest, similar to Section \ref{sec:Zevap_constraint}.

\subsection{Constraints from critical temperature}\label{sec:Tcrit_constraint}

This scenario also requires that the PBHs emit a blackbody spectrum above the threshold temperature to suppress $H_2$ formation and delay gas collapse to a massive atomic cooling halo ($T_{\rm crit} \sim 8000$ K) \citep{Sugimura_Omukai_Inoue_2014}. This temperature corresponds to photons of energy $E_\nu = 2.0$ eV, which are required for the photodissociation of $H^-$, an important element in the process of $H_2$ production. This sets an upper limit to the masses of PBHs of interest, i.e. $M_{\rm PBH} < M_{\rm crit}$. From eq. (\ref{eq:Temperature_HR}), we get a mass constraint: $M_{\rm crit} \lesssim 7.5 \times 10^{-12} \: \rm M_\odot$ (or $1.5 \times 10^{22}$ g).

\begin{figure}[h]
    \centering
    \includegraphics[width=0.75\linewidth]{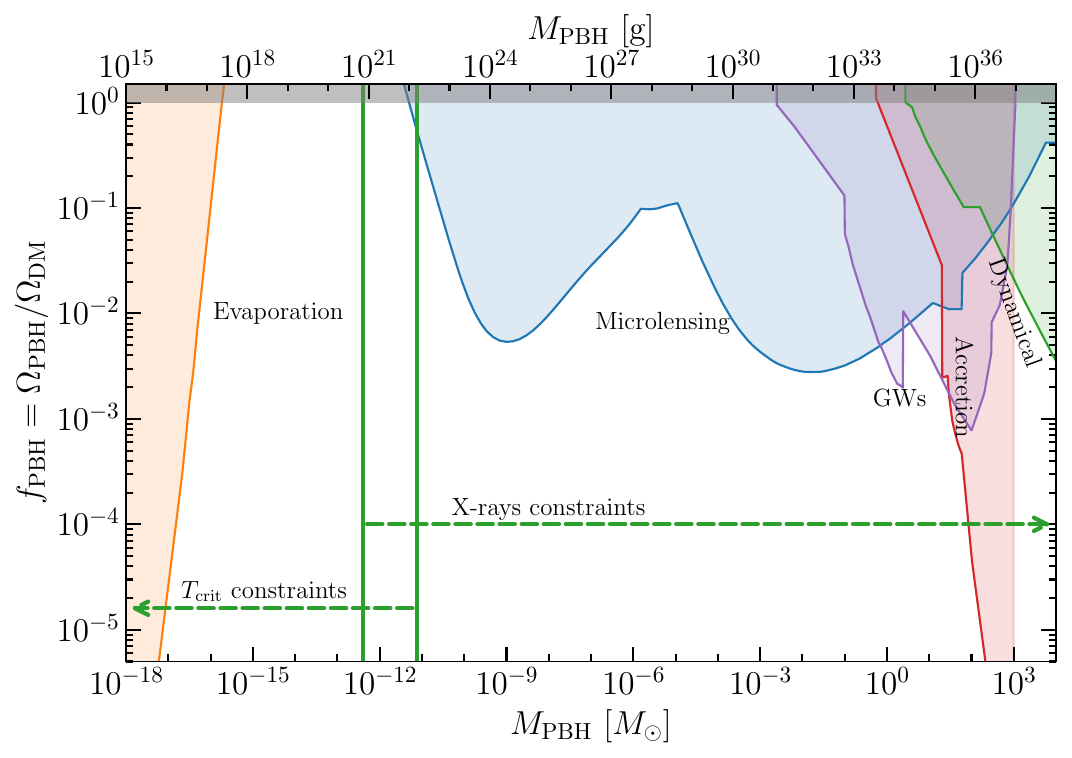}
    \caption{Constraints on the fraction of DM in PBHs from different methods (filled regions - Hawking evaporation (orange), gravitational lensing constraints on compact objects (blue), constraints from a stochastic background and individual gravitational wave mergers (purple), accretion on PBHs (red), dynamical constraints on compact objects (green)), based on \cite{Green_Kavanagh_2021} and plotted using \texttt{PBHbounds}\footnote{Which can be found [\href{https://github.com/bradkav/PBHbounds}{here}], with the official release [\href{https://zenodo.org/records/3538999}{here}]. The filled contour constraints are made from a combination of multiple individual constraints that belong to each category. For detailed references, we refer to \cite{Green_Kavanagh_2021}.}. The dashed lines show mass ranges that satisfy the respective constraints labeled above the arrows. These are physical constraints on the masses of PBHs that are needed to satisfy some of the DCBH formation criteria, as described in Sections \ref{sec:Xray_constraint} and \ref{sec:Tcrit_constraint}. The constraints based on evaporation redshift,  i.e. that the PBHs have not evaporated by $z=10$ - see Section \ref{sec:Zevap_constraint}, are not plotted here since the allowed region from that constraint spans the full plot. The y-axis values on the constraints are arbitrary and chosen for clearer visualization.}
    \label{fig:PBHbounds_DCBHs_Tz}
\end{figure}

The constraints reported in Sections \ref{sec:Zevap_constraint}, \ref{sec:Xray_constraint} and \ref{sec:Tcrit_constraint} are necessary but not sufficient conditions for PBHs to act as the appropriate radiation sources to decrease the cooling rates of the primordial gas clouds. These constraints simply establish the range of masses that are relevant for this scenario, which is graphically depicted by the dashed lines in Figure \ref{fig:PBHbounds_DCBHs_Tz}. Note that these mass limits do not pose any restrictions to $f_{\rm PBH}$; they simply point to the PBH masses that are required for this mechanism to be valid. Interestingly, the $T_{\rm crit}$ and X-ray constraints demand masses in a region of $f_{\rm PBH}$ parameter space that remains predominantly open: $4.0 \times 10^{-13} \: \rm M_\odot \lesssim M_{\rm PBH} \lesssim 7.5 \times 10^{-12} \: \rm M_\odot$. 

\begin{figure}
    \centering
    \includegraphics[width=0.48\textwidth]{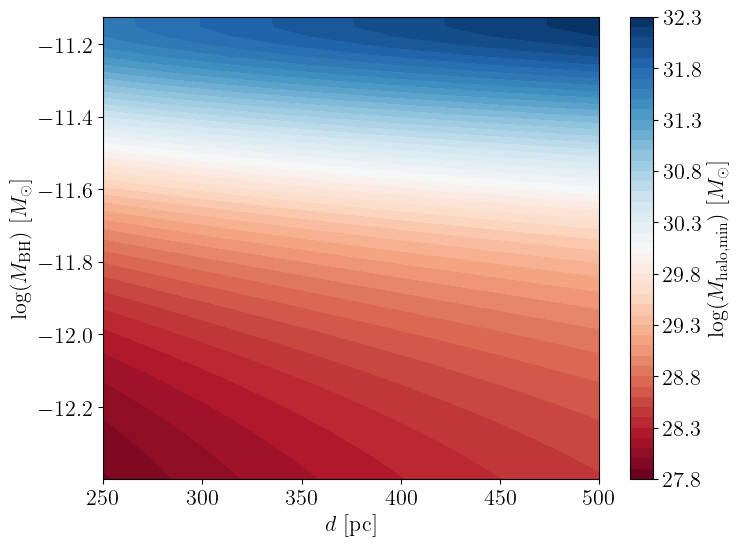}
    \includegraphics[width=0.49\textwidth]{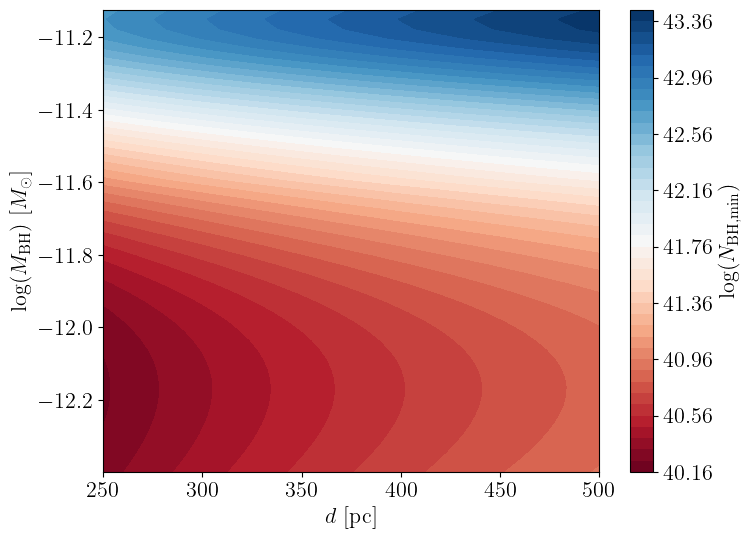}
\caption{{\bf(a)} 
Minimum estimated PBH halo mass (as a function of PBH mass and distance) that is required to produce $J_{21, \rm LW}^{\rm tot} \gtrsim J_{\rm crit}(M_{\rm PBH})$
for the simplest scenario considered in this work: a population of monochromatic PBHs within the mass range of interest at a distance $d$ from a primordial gas cloud. The minimum mass of the PBH halos that satisfy this condition ($M_{\rm h} \sim 10^{27.8} \rm M_\odot$) are many orders of magnitudes larger than the most massive DM halos expected, rendering this scenario unphysical. 
{\bf (b)} Same as (a), but parameterized in terms of the minimum number of PBHs that are required in order to reach $J_{21, \rm LW}^{\rm tot} \gtrsim J_{\rm crit}(M_{\rm PBH})$. For the distances, we adopt the range [$250, 500$] pc, to compare with \cite{Visbal_et_al_2014}. This range ensures the halos to not be tidally disrupted by their interactions, but be close enough for the LW intensity from stars to be significant. See section \ref{sec:ConstDist_Monochromatic} for more details.}
\label{fig:Constant_D_monochromatic}
\end{figure}

\newpage
\subsection{Constraints from critical radiation intensity}\label{sec:CRI}

In this section, we investigate whether it is feasible for PBHs within the mass range of interest (determined in Sections \ref{sec:Zevap_constraint}, \ref{sec:Xray_constraint} and \ref{sec:Tcrit_constraint}) to provide the critical radiation intensity (CRI) $J_{\rm crit}$ needed to prevent the collapse of the initial cloud, in order to happen in a massive, atomic-cooling halo to achieve high accretion rates. CRI is given in units of erg/s/cm$^2$/Hz/sr, and is commonly normalized with respect to $J_{21} = 10^{-21}$ erg/s/cm$^2$/Hz/sr \citep{Bromm_Loeb_2003, Loeb_Furlanetto_2013, Sugimura_Omukai_Inoue_2014}.

Across the literature, there are a number of different estimates for the CRI needed to keep the primordial cloud heated \cite[e.g.][]{Agarwal_Khochfar_2015, Agarwal_et_al_2016, Agarwal_et_al_2016_CR7, Agarwal_et_al_2017, Latif_Khochfar_2019_uv, Agarwal_et_al_2019_cooling}. These estimates depend on several key features such as the relevant radiation source and its spectrum, the spatial configuration of the halo, and the cosmological setting, with $J_{\rm crit}$ ranging from $\sim1$ to $10^5$ \citep[e.g.][]{Omukai_2001, Bromm_Loeb_2003, Shang_Bryan_Haiman_2010, Visbal_et_al_2014, Agarwal_Khochfar_2015, Agarwal_et_al_2016, Latif_Ferrara_2016, Wolcott-Green_et_al_2017, Incatasciato_et_al_2023}. In our approach, we follow \cite{Sugimura_Omukai_Inoue_2014}, and adopt a $J_{\rm crit}$ that depends on the radiation temperature. More specifically, \cite{Sugimura_Omukai_Inoue_2014} provides a fitting formula for $J_{\rm crit}$ with respect to the ratio of photodissociation coefficients of $H^-, H_2$, i.e. $J_{\rm crit} = f(k_{H^-}/k_{H_2})$, which in turn depends on the temperature of the blackbody spectrum. In our case, we connect the temperature of the HR to the masses of the PBHs, and have a mass-depended critical intensity $J_{\rm crit} = f(M_{\rm PBH})$. We note that their fitting formula is valid for a temperature range $T\ \epsilon\ (8000, 2\cdot 10^5)$ K, which is consistent with our temperature constraints in Section \ref{sec:Tcrit_constraint}. For our range of masses, the minimum and maximum $J_{\rm crit}$ are $5.12$ and $1400$ respectively in the standard $J_{21}$ units.

We note that recent studies \citep{Schauer_et_al_2021, Kulkarni_Visbal_Bryan_2021, Munoz_et_al_2022} also show that the presence of streaming velocities generally compliments the effects of LW radiation, allowing heavy seed formation even under moderate fragmentation. The relative importance of the two mechanisms - streaming velocities and LW radiation intensity - depends on redshift, with the former dominating at early times and the latter dominating at redshifts $z \lesssim 15$. Although these works analyze the critical (minimum) mass of dark matter halos required for PopIII star formation, compared to the status of gas clouds investigated in our case, the CRI found is only approximately one order of magnitude smaller than the values we consider, so it does not change the conclusions presented here.

To provide an easily-reproducible analysis framework for the scientific community, we follow \cite{Liu_Zhang_Bromm_2022, Liu_Bromm_2023} by treating the LW radiation from the PBHs as a `background' and ignoring self-shielding -- the idea that a fraction of the outer layer of the gas protects the inner layer from radiation, allowing it to cool.

A BH with Schwarzschild radius $R_{\rm S}$ emits HR with luminosity $L_{\rm LW} = 4 \pi^2 R_{\rm S}^2 B_{\rm LW}$. Its flux at distance $d$ and solid angle $\Omega = 4 \pi$ sr, assuming spherical geometry, give a specific intensity:

\begin{equation}\label{eq:J_LW_single_BH}
    J_{\rm LW} = \frac{B_{\rm LW}}{4} \left( \frac{R_{\rm S}}{d}\right)^2.
\end{equation}

In standard units, this becomes:

\begin{equation}
    J_{21, \rm LW} = 10^{21}\frac{B_{\rm LW}}{4} \left( \frac{R_{\rm S}}{d}\right)^2.
\end{equation}

In the following subsections, we investigate several different PBH spatial distributions. Unless stated otherwise, we calculate PBH densities assuming the most optimistic limit of $f_{\rm PBH}=1$, which is allowed for the mass range found in Sections \ref{sec:Zevap_constraint}, \ref{sec:Xray_constraint} and \ref{sec:Tcrit_constraint}. Note that we do not present results for different mass distributions because we show in Appendix \ref{sec:Ap_mass_functions} that a population of monochromatic PBHs is sufficient to test the viability of the scenario at hand.

\subsubsection{Constant distance \& monochromatic PBHs}\label{sec:ConstDist_Monochromatic}

The simplest case, as explored in \cite{Visbal_et_al_2014, Wise_et_al_2019}, is a cluster of monochromatic PBHs at a distance $d$ from a primordial gas cloud that irradiates the cloud with LW HR photons\footnote{For simplicity, we focus on photons in the middle of the energy range, i.e. we take $E_\gamma = 12.5$ eV. Alternatively, one can integrate $\int_{\nu_1}^{\nu_2} L_\nu d\nu$ and divide by $\Delta\nu$.}. For $N_{\rm BH}$ black holes of the same mass, the total intensity is $J_{21, \rm LW}^{\rm tot} = N_{\rm BH} \cdot J_{21, \rm LW}$. We find $N_{\rm BH}$, by demanding that $J_{21, \rm LW}^{\rm tot}>J_{\rm crit}(M_{\rm PBH})$. 
Figure \ref{fig:Constant_D_monochromatic}(a) shows an approximation of the total PBH halo mass ($M_{\rm PBH, halo} = N_{\rm BH} M_{\rm BH}$) as a function of distance $d$ for the range of allowed PBH masses. 
Even for the most optimistic combinations of $d$ and $M_{\rm BH}$, the total halo mass required for this scenario ($M_{\rm h} \sim 10^{27.8} \rm M_\odot$) exceeds the expected halo mass function at $z>10$ by several orders of magnitude \cite[e.g.][]{Lovell_et_al_2023, OBrennan_Regan_Power_2024}, indicating that such a configuration is not valid for the PBHs of interest here. Of course, this remains the case even if the lower limit of $J_{21, \rm LW}$ is used. Figure \ref{fig:Constant_D_monochromatic}(b) shows the actual number of PBHs required for a specific combination of mass and distance.

Note that here we do not consider the DCBH-formation timescale. In contrast, \cite{Visbal_et_al_2014} considers a timescale of $t_{\rm coll} \sim 10$ Myr. Both approaches are valid given the different mechanisms in the two situations. In \citep{Visbal_et_al_2014}, they consider that the LW radiation is provided by the second member of a pair of atomic cooling halos, which manages to form stars first. The latter are the source of the LW photodissociating radiation and evolve in similar timescales as the collapse. In our case, as shown in Appendix \ref{sec:Ap_PBHs_evolution_redshift}, the time evolution of PBHs is negligible for the redshift and mass ranges of interest. In other words, emission is expected to be constant for the whole period of collapse by default.

\subsubsection{Uniform distribution \& monochromatic PBHs}\label{sec:Uniform_Monochromatic}

Next, we consider a population of monochromatic PBHs that are uniformly distributed inside the halo from a minimum radius $r_{\rm min}$ to a maximum radius $r_{\rm max}$. This maximum radius is based on an estimate of the halo size, as explained in Appendix \ref{sec:Ap_Halos_properties}. The choice of minimum value is arbitrary, but motivated from \cite{Liu_Zhang_Bromm_2022}, who find that stellar-mass PBHs can reach distances on the order of $\sim 1$ pc from the center of the DM halo. We choose $r_{\rm min} = 1$ pc, unless stated otherwise, to be able to compare between the two scenarios.

The density of a uniform-density shell of volume $V_{\rm shell} = 4 \pi (r_{\rm max}^3 - r_{\rm min}^3)/3$ is given by:
\begin{equation}
     \rho = \rho_{\rm vir} = \rho_0 = \frac{M_{\rm tot}}{V_{\rm shell}} = \frac{N_{\rm BH} M_{\rm BH}}{V_{\rm shell}}.
\end{equation}
The number of PBHs within the shell is then given by:
\begin{equation}
    N_{\rm BH} = \frac{\rho_0}{M_{\rm BH}} \cdot \frac{4}{3} \pi (r_{\rm max}^3 - r_{\rm min}^3),
\end{equation}
or for a thin shell, the differential number of PBHs is:
\begin{equation}
    dN_{\rm BH} = \frac{\rho_0}{M_{\rm BH}} \cdot 4 \pi r^2 dr.
\end{equation}

Then for the whole volume, the total LW radiation intensity at the center of the halo is:
\begin{equation}\label{eq:J21_uniform}
    J_{21, \rm LW}^{\rm tot} = \int_{r_{\rm min}}^{r_{\rm max}} dN_{\rm BH} J_{21, \rm LW}  = 10^{21} \frac{\pi B_{\rm LW} R_{\rm S}^2\rho_0}{M_{\rm BH}} (r_{\rm max}-r_{\rm min}) = 10^{21} \frac{\pi B_{\rm LW} R_{\rm S}^2 \rho_0 r_{\rm min}}{M_{\rm BH}} \left(\frac{r_{\rm max}}{r_{\rm min}} -1 \right).
\end{equation}

\begin{figure}[h]
    \centering
    \includegraphics[width=1.0\linewidth]{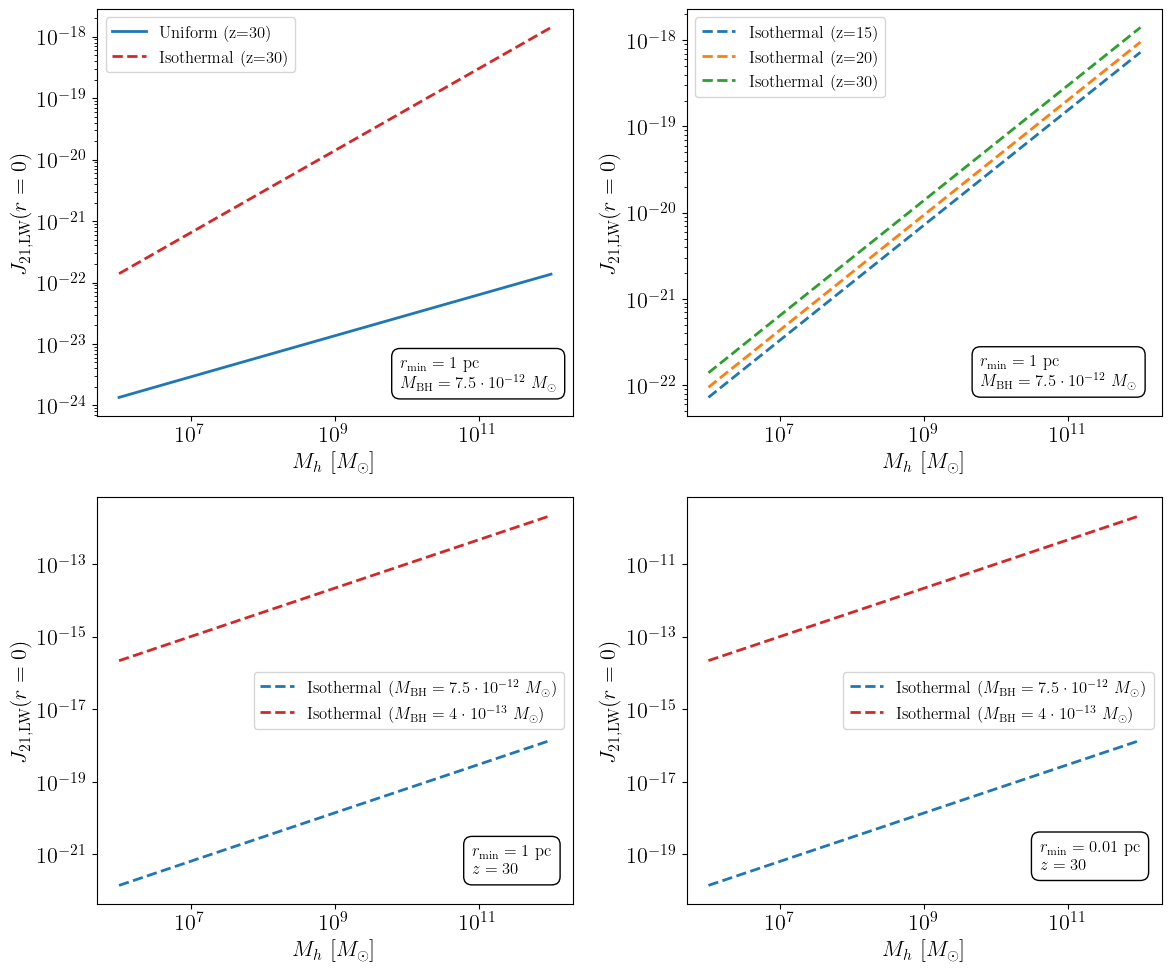}
    \caption{Specific intensity at the center of the halo for different halo masses. PBHs are distributed within the halo based on a uniform or isothermal density (Section \ref{sec:Uniform_Monochromatic} versus Section \ref{sec:Isothermal_Monochromatic}). The four panels of this figure investigate the effects of different density profiles (top left), of different initial redshifts of collapse (top right), of different PBH masses (bottom left), and of different minimum distances that the PBHs can reach inside the halo (bottom right). The density at $r_{\rm max}$ is set to the virial density $\rho_0 = \rho_{\rm vir}$, as described in Appendix \ref{sec:Ap_Halos_properties}. The emitted specific intensity for all cases that we investigate remains orders of magnitude below the CRI required for collapse ($5.12 \lesssim J_{\rm crit} \lesssim 1400$), indicating that HR from non-evaporating PBHs is not in itself a sufficient mechanism for the formation of DCBHs in the early Universe.}
    \label{fig:Jcrit_Density_monochromatic}
\end{figure}

The blue line in the top left panel of Figure \ref{fig:Jcrit_Density_monochromatic} shows the specific intensity of LW photons at the center of a halo as a function of total halo mass. Even for the largest halo masses in this case, the radiation intensity is many orders of magnitude lower than the CRI needed to form a DCBH. Note that varying the minimum radius, redshift, and PBH mass affects the total radiation intensity slightly -- as explored in the following subsection for the isothermal case -- but not enough to reach the required $J_{\rm crit}$. As with the case of monochromatic PBHs at a constant distance, a uniform distribution of monochromatic, non-evaporating PBHs similarly cannot enable the collapse of a DCBH. 

\subsubsection{Isothermal distribution \& monochromatic PBHs}\label{sec:Isothermal_Monochromatic}

For a more realistic case, we investigate a monochromatic population of PBHs that follow an isothermal distribution \cite[e.g.][]{Wise_et_al_2019, Liu_Zhang_Bromm_2022}. Here, the density scales as $\rho = \rho_0 (r_{\rm max}/r)^2$, where the density equals the virial density of the halo for $r=r_{\max}$.

In this case, the total LW radiation intensity from the shell at the center of the halo is:
\begin{equation}\label{eq:J21_isothermal}
    J_{21, \rm LW}^{\rm tot} = \int_{r_{\rm min}}^{r_{\rm max}} dN_{\rm BH} J_{21, \rm LW}  = 10^{21} \frac{\pi B_{\rm LW} R_{\rm S}^2 \rho_0 r_{\rm max}^2}{M_{\rm BH}} \int_{r_{\rm min}}^{r_{\rm max}} \frac{dr}{r^2} = 10^{21} \frac{\pi B_{\rm LW} R_{\rm S}^2 \rho_0 r_{\rm max}}{M_{\rm BH}} \left(\frac{r_{\rm max}}{r_{\rm min}} -1 \right).
\end{equation}

Comparing this case with eq. (\ref{eq:J21_uniform}) gives:
\begin{equation}
 J_{21}^{\rm isothermal} = (r_{\rm max}/r_{\rm min}) J_{21}^{\rm uniform}.   
\end{equation}

Since $r_{\max}$ > $r_{\min}$ and thus $J_{21}^{\rm isothermal} > J_{21}^{\rm uniform}$, an isothermal distribution of PBHs will provide a specific intensity of LW photons that is {\it closer} to the required $J_{\rm crit}$ value than a uniform distribution of PBHs will, as demonstrated in the top left panel of Figure \ref{fig:Jcrit_Density_monochromatic}. However, even for the largest halos, the specific intensity still falls several orders of magnitude short of the required $J_{\rm crit}$ for this scenario. 

Figure \ref{fig:Jcrit_Density_monochromatic} also shows the effects of varying other parameters such as the minimum distance of the PBHs from the center, the formation redshift, and the PBH masses within the allowed ranges. None of these cases provide a radiation intensity that reaches the required $J_{\rm crit}$ value, even in the more optimistic scenario where $J_{\rm crit} \sim 5$. We find that to reach $J_{\rm crit} \sim 5$, in redshift $z \sim 30$ for a halo of mass $M_{\rm h} \sim 10^{-13} \: \rm M_\odot$ -- the more optimistic scenario -- would require an $r_{\rm min} \sim 3$ km. This is unphysical, if one considers that the Schwarzschild radius of a BH with $M_{\rm BH} \sim 10^5 \: \rm M_\odot$ (which would be the approximate mass of the seed BH at the center) would be much bigger than that.

Hence, the investigated radiation mechanism provides a LW intensity many orders of magnitude smaller than critical, yielding it an ineffective way of creating the seeds of massive BHs in the early Universe.

\subsubsection{LW background from monochromatic PBHs}

Finally, we consider the effects of the PBHs' LW radiation acting as a cosmological background \citep{Haiman_et_al_1997, Visbal_et_al_2014_LWB}. To calculate the LW background, we take a similar approach to the one of the background produced by stars \citep{Schauer_Liu_Bromm_2019} and use:

\begin{equation}\label{eq:LW_background}
    J_{\rm LW, 21}^{\rm bg}(z) = 10^{21}\frac{c}{4 \pi} \int_z^{z_{\rm max}} dz' \left|\frac{dt}{dz'}\right| \left( \frac{1+z}{1+z'} \right)^3 j_{\rm LW}(z', z),
\end{equation}

where $dt/dz'$ denotes the cosmic time evolution with redshift, and $j_{\rm LW}$ gives the proper specific intensity defined as

\begin{equation}
    j_{\rm LW} (z', z) = L_{\rm LW} (z', z) n(z),
\end{equation}

where $L_{\rm LW}$ is the luminosity of the sources and $n$ is their number density. The redshift $z$ denotes the time when we estimate the background radiation, while $z'$ covers the redshift range on which a different blackbody photon would be redshifted to the LW band at $z$, i.e. $\nu' = \nu_{\rm LW} (1+z')/(1+z)$. The maximum redshift is estimated through the `screening approximation', which assumes that the LW photons can travel through the intergalactic medium until they get absorbed by a Lyman line:

\begin{equation}
    \frac{1+z_{\rm max}}{1+z} = \frac{\nu_{\rm max}}{\nu} = f.
\end{equation}

Based on \cite{Visbal_et_al_2014}, we assume $f \sim 1.04$.

Analysis of the LW background from stars in cosmological simulations provides an estimate of $J_{\rm LW, 21}^{\rm bg} \sim 3.6$ \citep{Incatasciato_et_al_2023}. We compare this with the case of a population of monochromatic PBHs. The proper specific intensity becomes:

\begin{align}\label{eq:LW_proper_spec_int}
    j_{\rm LW} (z) &= L_{\rm LW} (z', z) n_{\rm PBH}(z) \nonumber \\ 
    &= \frac{L_{\rm LW}}{M_{\rm BH}} M_{\rm BH} n_{\rm PBH}(z) \nonumber \\ 
    & = \frac{L_{\rm LW}}{M_{\rm BH}} f_{\rm PBH} \rho_{\rm DM}(z) \nonumber \\ 
    &= \frac{L_{\rm LW}}{M_{\rm BH}} f_{\rm PBH} \Omega_m(z) \rho_{\rm crit}(z) \nonumber \\ 
    &= \frac{4 \pi^2 R_{\rm S}^2 B_{\rm LW}(z', z)}{M_{\rm BH}} f_{\rm PBH} \Omega_{m, 0} (1+z)^3 \frac{3 H_0^2}{8 \pi G},
\end{align}

with

\begin{equation}
    B_{\rm LW}(z', z) = B_{\rm LW} \left( \nu_{\rm LW} \frac{1+z'}{1+z} \right),
\end{equation}

and where we have connected the number density of PBHs to that of DM, and used the cosmological quantities defined in Appendix \ref{sec:Ap_Halos_properties}. Replacing $|dt/dz|$ with $1/(1+z)H(z)$ and substituting the expression for the proper specific intensity from eq. (\ref{eq:LW_proper_spec_int}) into eq. (\ref{eq:LW_background}) yields:

\begin{equation}
    J_{\rm LW, 21}^{\rm bg} = 10^{21}\frac{c}{4 \pi}  \frac{4 \pi^2 R_{\rm S}^2}{M_{\rm BH}} f_{\rm PBH} \Omega_{m, 0} \frac{3 H_0^2}{8 \pi G} \int_z^{z_{\rm max}} dz' \left( \frac{1+z}{1+z'} \right)^3 \frac{B_{\rm LW}(z', z)(1+z')^3}{(1+z')H(z')},
\end{equation}

which simplifies to

\begin{equation}
    J_{\rm LW, 21}^{\rm bg} = 10^{21} \frac{c R_{\rm S}^2}{M_{\rm BH}} f_{\rm PBH} \Omega_{m, 0} \frac{3 H_0^2}{8 G} (1+z)^3 \int_z^{z_{\rm max}} dz' \frac{B_{\rm LW}(z', z)}{(1+z')H(z')}.
\end{equation}

For a quick estimate of its value, we observe that

\begin{equation}
    \int_z^{z_{\rm max}} dz' \frac{B_{\rm LW}(z', z)}{(1+z')H(z')} \sim \frac{B_{\rm LW}(\nu_{\rm LW})}{(1+z)H(z)}\Delta z,
\end{equation}

where, as a reminder, we set $\nu_{\rm LW}$ to the photons with energy in the middle of the LW band ($E_{\rm LW} = 12.5$ eV) and have

\begin{equation}
    \Delta z = z_{\rm max}-z = (1+z_{\rm max})-(1+z) = (f-1)(1+z).
\end{equation}

Finally, this leads to 

\begin{equation}\label{eq:LW_bg_approx}
    J_{\rm LW, 21}^{\rm bg} \sim 10^{21} \frac{c R_{\rm S}^2 B_{\rm LW}}{M_{\rm BH}} f_{\rm PBH} \Omega_{m, 0} \frac{3H_0^2}{8 G} (1+z)^3 \frac{(f-1)}{H(z)}.
\end{equation}

For $z=10$ and $M_{\rm PBH}=4 \cdot 10^{-13} - 7.5 \cdot 10^{-12} M_\odot$, this results in $J_{\rm LW, 21}^{\rm bg}\sim 10^{-26}-10^{-20}$, which is extremely small, and much smaller than the stellar contribution. Note that this more optimistic scenario is equivalent to assuming a flat spectrum in the LW band, and gives an upper limit. Therefore, the LW background radiation from PBHs does not play any role in blocking H$_2$ formation.

\section{Discussion}\label{sec:Discussion}

This section discusses the simplifying assumptions made in this paper, and assesses their influence on the results. 

\subsection{What happens with the production of other particles from HR?}\label{sec:massless_particles}

In this paper, we have assumed that HR from PBHs consists only of massless particles -- specifically, photons. In reality, HR can produce a variety of massless and massive particles \citep{Page_1976, Frolov_Novikov_1998}. These will be important for both the radiation signature of the black hole and its evolution. We comment on these impacts in order below, but first provide a simple estimate of when these effects are important.

As mentioned in Section \ref{sec:Hawking_radiation}, the temperature of HR is inversely proportional to the mass of a BH. As a result, as the BH shrinks it becomes more energetic and could have enough energy to emit massive particles. Ignoring neutrinos and taking the electron as the smallest nonzero-rest-mass particle, massless particles will dominate HR when:

\begin{equation}
    k_B T \lesssim m_e c^2 \Rightarrow M_{\rm BH} \gtrsim \frac{\hbar c}{8 \pi G m_e} \Rightarrow M_{\rm BH} \gtrsim 6.5 \times 10^{-17}\ M_\odot \sim 1.3\times 10^{17}\ {\rm g}.
\end{equation}

When the BH masses are above this threshold, as is the case for the PBHs of interest in this work (see Section \ref{sec:PBHs_constraints}), considering only the primary spectra is a valid approximation. This regime is denoted by the region where $m_{\rm HR} \rightarrow 0$ in Figure \ref{fig:PBHbounds_DCBHs_mGrey}.

When the BH masses are below this threshold, massive particles start being produced\footnote{This is a conservative threshold, since electrons and positrons would be ultra-relativistic at the range $m_e c^2 \ll k_B T \ll m_\mu c^2$, where $m_\mu$ is the mass of the muon. Taking this inequality into account by checking $k_B T \lesssim m_\mu c^2$, \cite{Page_1976} estimates that the massless particles approximation will break down for masses $M_{\rm BH} \lesssim 10^{14}$ g.}. These particles are not the final end products for an energetic BH since they can decay further \citep{BlackHawk_2019, Lu_Picker_Kusenko_2024}. A percentage of these secondary particles would lead to the emission of additional photons -- a `secondary spectrum'. 

The effects of this extra emission are two-fold:

\begin{enumerate}
    \item  It will increase the emitted flux for a given BH. For the scenario investigated in this paper, the secondary spectra would strengthen the impact of HR by PBHs (see, for example, Figures 4 and 3 in \cite{BlackHawk_2021} and \cite{Auffinger_2023} respectively). In principle, focusing only on the primary spectra makes our analysis a conservative approach. However, the impact is negligible here since the mass regime of importance to our analysis (based on the constraints in Section \ref{sec:PBHs_constraints}) is greater than the mass threshold where the secondary spectra become significant.

    \item It will lead to an increased rate of mass loss for a BH, expediting its evaporation \citep{Mosbech_Picker_2022}.

The mass evolution in this case is given by \citep{Mosbech_Picker_2022}:

\begin{equation}\label{eq:effective_mass_evolution}
    M(t) = \left(M_i^3 - 3 \alpha_{\rm eff} \frac{\hbar c^4}{G^2}t \right)^{1/3},
\end{equation}

where $\alpha_{\rm eff}=f(M)$ is a function of mass for $M_i \lesssim 10^{18}\ {\rm g}$, or a constant for $M_i \gtrsim 10^{18}\ {\rm g}$. For the `standard' case we consider in this paper, $\alpha_{\rm eff} = 1/15360\pi$. Although this would affect our determination of the evaporation time of a PBH, in the redshift and mass ranges we are considering, the PBHs are significantly far from evaporation and their evolution is effectively negligible (Figure \ref{fig:mass_evolution_redshift}). Therefore, this will not significantly affect our results.

\end{enumerate}

\subsection{Where are the `Greybody factors'?}\label{sec:greybody_factors}

In the original calculation of Hawking radiation \citep{Hawking_1974, Traschen_2000_Intro_HR} the number of emitted particles is

\begin{equation}
    N_\nu = \Gamma_\nu \frac{1}{e^{h \nu/k_B T}-1},
\end{equation}

where the $\Gamma_\nu$ term signifies the `greybody' factor that modifies the spectrum from a perfect blackbody. It acts as the absorption coefficient for scattering of a scalar field off the gravitational field of a BH \citep{Traschen_2000_Intro_HR, Carroll_2019}. In the short-wavelength/high-frequency limit where a photon has enough energy to surpass the effective potential due to the spacetime of the BH, this backscattering effect is small and can be ignored. To provide a quick estimate of where this limit occurs, we compare the wavelengths of the photons produced by HR with the Schwarzschild radius of the BH. 

For the `greybody' factors to {\it not} play an important role, the wavelength must be smaller than the Schwarzschild radius:

\begin{equation}
    \lambda \lesssim R_{\rm S} = \frac{2 G M}{c^2} \Rightarrow M \gtrsim \frac{hc^3}{2GE}\quad {\rm or}\quad \nu \gtrsim \frac{c^3}{2GE},
\end{equation}

where $E=h \nu$ is the energy of the radiated photon, and $c=\lambda \nu$ is the speed of light. Setting $E=E_{\rm LW} \sim 12.5$ eV for the LW energy band provides the black hole mass threshold above which the `greybody' factors can be ignored:

\begin{equation}
    M_{\rm BH} \gtrsim 3.35 \times 10^{-11}\ M_\odot \sim 6.68\times 10^{22}\ {\rm g}.
\end{equation}

\begin{figure}[h]
    \centering
    \includegraphics[width=0.75\linewidth]{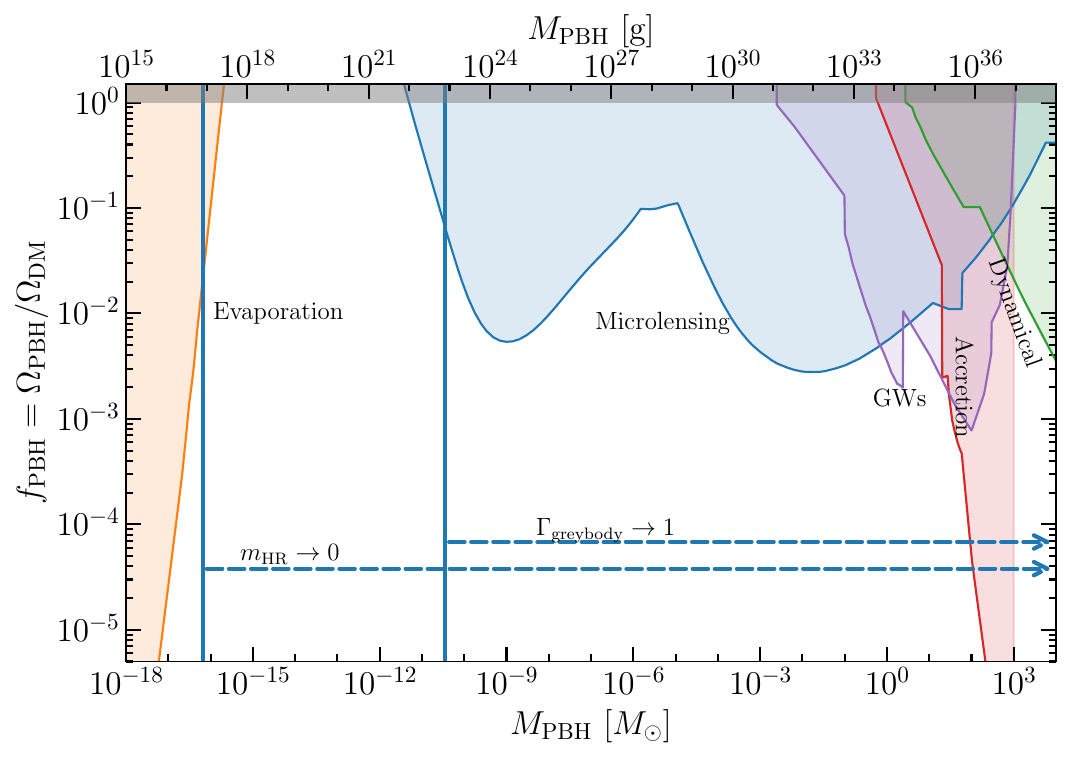}
    \caption{Similar to Figure \ref{fig:PBHbounds_DCBHs_Tz}, but now the dashed lines show mass ranges (specified by the arrows) where various assumptions are valid.
    $m_{\rm HR} \rightarrow 0$ indicates the region where it is valid to only consider the primary HR spectra. The PBH mass range considered for this work {\it does} fall within this region of parameter space, and as such, we safely adhere to this assumption. 
    $\Gamma_{\rm greybody} \rightarrow 1$ indicates the region where it is valid to ignore greybody factors when calculating the HR spectrum. The PBH mass range considered for this work does {\it not} fall within this region of parameter space, and as such, we do account for a greybody factor. 
    The y-axis values chosen for these are arbitrary and for visualization purposes only. For more details on these limits, see Sections \ref{sec:massless_particles} and \ref{sec:greybody_factors}.}
    \label{fig:PBHbounds_DCBHs_mGrey}
\end{figure}

The regime where `greybody' factors may be ignored is denoted by the region where $\Gamma_{\rm greybody} \rightarrow 1$ in Figure \ref{fig:PBHbounds_DCBHs_mGrey}. The mass range of the PBHs of interest in this work (see Section \ref{sec:PBHs_constraints}), however, lie in this regime where $\Gamma \neq 1$. As such, we do need to take `greybody' factors into account. In principle, codes calculating Hawking radiation deal with `greybody` factors in a systematic way by also monitoring all of the particles that are produced (see Section \ref{sec:massless_particles}). For simplicity, we instead multiply the spectrum by a constant `greybody' factor, $f_{\Gamma}=0.24$, based on values in Section IV and Table I of \cite{Page_1976}. This constant factor corresponds to the ratio of photon power calculated numerically with the inclusion of the necessary `greybody' factors to the photon power in the high frequency/energy limit.

\subsection{Comparison to previous works}

As shown in Section \ref{sec:CRI}, none of the scenarios considered in this work contributed enough radiation to reach the critical LW threshold. This may seem in conflict with the recent results of \cite{Lu_Picker_Kusenko_2024} where, under certain circumstances, PBHs were able to provide sufficient HR intensity to allow for direct collapse of the cloud. However, in reality, our work is complementary and consistent with their conclusions in the following ways:

\begin{itemize}
    \item One main difference is that we are not considering evaporating BHs ($M_{\rm BH} \sim 10^{14}$ g). The final `explosions' of evaporating black holes produce high-energy radiation with an enhanced intensity due to the presence of significant secondary spectra. This explains the apparent disagreement with our work, which deals only with primary spectra and with masses of PBHs a few orders of magnitude bigger (and hence less energetic). Note that both mass ranges are viable; however, for our mass range, observational constraints allow $f_{\rm PBH} = 1$, while for smaller PBHs ($M_{\rm BH} \lesssim 10^{14}$ g) there are severe constraints leading to $f_{\rm PBH} < 10^{-10}$ for the global fraction of PBHs. 

    \item In our case, we are considering the intensity of HR from PBHs compared to the critical LW radiation required for the dissociation of $H_2$, and other species connected to $H_2$ formation. The lack of these molecules lead to a reduced cooling rate, and hence the accumulation of bigger gas mass. This process is similar, but conceptually different from \cite{Lu_Picker_Kusenko_2024} who consider the direct deposit of energy in the gas cloud.
    
    \item \cite{Lu_Picker_Kusenko_2024} finds that significant PBH clustering is necessary inside the halos -- about 7 orders of magnitude bigger than the external density, i.e. $f_{\rm PBH, in} \sim 10^7 f_{\rm PBH, out}$ -- in order for collapse to a DCBH to occur. Although the models that we investigate do not produce radiation with sufficient intensity to surpass the CRI, the fact that the isothermal distribution of PBHs gets closer to reaching the required CRI than does the uniform distribution of PBHs highlights -- at least qualitatively -- the importance of PBH density/clustering. Further analysis is required to determine if there are any extreme spatial distributions of PBHs within our mass range that could enable the collapse of a DCBH.
\end{itemize}

\section{Conclusions}

The formation of supermassive black holes at the center of distant galaxies is an open question. Direct collapse black holes offer a solution. However, they require specific conditions in a primordial gas cloud in order to be formed -- namely, the gas must remain warm enough, through a reduced cooling rate, and hence avoid fast and significant fragmentation to smaller clumps. In this work, we considered Hawking radiation from non-evaporating primordial black holes as the mechanism to destroy $H_2$ and related coolant species.

We used a simplified model of photon emission due to Hawking radiation from Schwarzschild black holes and calculated their specific intensity in the Lyman-Werner band, which plays a critical role in this scenario. We used physical constraints based on the evaporation timescale, X-ray emission, and temperature of primordial black holes to estimate a range of masses that is feasible for this scenario: $M_{\rm PBH}\ \epsilon \ (4 \cdot 10^{-13}, 7.5 \cdot 10^{-12}) \ \rm M_\odot$. Interestingly, all values for the fraction of dark matter contained in primordial black holes are still allowed by observational constraints for this range of PBH masses. For PBHs within this mass range, we evaluated whether different spatial distributions of PBHs in or near a primordial gas cloud could produce the critical radiation intensity needed for the formation of a massive, atomic-cooling cloud, enabling it to directly collapse into a heavy seed for the early-Universe SMBHs that have been observed. 

We conclude that non-evaporating PBHs cannot provide the radiation needed to facilitate the formation of massive black hole seeds since their radiation intensity is many orders of magnitude smaller than the expected critical value $J_{\rm crit} \sim 5 - 1400$. Nevertheless, we exclude a range of masses of PBHs that could potentially have acted as `LW sources'. Additionally, we qualitatively confirm the importance of significant clustering of PBHs, recently proposed in the literature, when considering the effects of Hawking radiation from PBHs in enabling the fomration of massive, atomic-cooling clouds.

\newpage
\section{Acknowledgments}

We would like to thank the referee for the very useful feedback that improved the clarity and quality of our paper. We also wish to thank the authors of \cite{Lu_Picker_Kusenko_2024} for useful clarifications and feedback on our work. Additionally, we thank Boyan Liu for helpful comments on the PBH distributions and the analysis of LW radiation in \cite{Liu_Zhang_Bromm_2022}, Sadegh Khochfar for bringing to our attention the X-ray constraints and providing helpful comments and references, and Charalampos Tzerefos for comments and useful references on PBHs.

\section{Data Availability}

We make our code public in \href{https://github.com/MariosNT/DCBHs_HR_PBHs}{\texttt{DCBHs\_HR\_PBHs}}.

\section{Appendix}

\subsection{Estimating halos sizes and masses}\label{sec:Ap_Halos_properties}

We assume a simple model for halo formation, where the mass of the collapsing DM halo is given by \cite{Cimatti_2020}:
\begin{equation}
    M_{\rm vir} = \frac{4 \pi}{3} \Delta_c \rho_{\rm crit} r_{\rm vir}^3,
\end{equation}
where $\rho_{\rm crit}$ is the critical density to make the Universe flat:
\begin{equation}
    \rho_{\rm crit} = \frac{3 H(z)^2}{8 \pi G},
\end{equation}
with $H(z)^2 = H_0^2 [\Omega_m (1+z)^3 + 1-\Omega_m]$. The $\Delta_c = \rho_{\rm vir}/\rho_{\rm crit}$ depicts the overdensity of the halo and depends on cosmology and redshift. Using the fit provided by \cite{Bryan_Norman_1998} we have:
\begin{equation}
    \Delta_c(z) = 18 \pi^2 + 82 y - 39 y^2,
\end{equation}
with $y = \Omega_m(z)-1$ and $\Omega_m(z)$:
\begin{equation}
    \Omega_m(z) = \frac{\Omega_{m, 0} (1+z)^3}{\Omega_m (1+z)^3 + 1-\Omega_m}.
\end{equation}
Solving for the virial radius, yields:
\begin{equation}\label{eq:virial_radius}
    r_{\rm vir} = \left(\frac{2 G M_{\rm vir}}{\Delta_c(z) H(z)^2} \right)^{1/3}.
\end{equation}
For a given redshift and a fixed halo mass, we use eq. (\ref{eq:virial_radius}) to get an estimate of the halo size.

Finally, the average density of the halo is given by $\rho_{\rm vir} = \Delta_c \rho_{\rm crit}$.

\begin{figure}[h]
    \centering
    \includegraphics[width=0.45\textwidth]{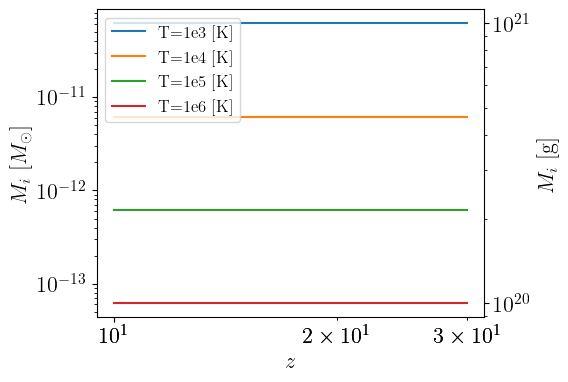}
    \includegraphics[width=0.45\textwidth]{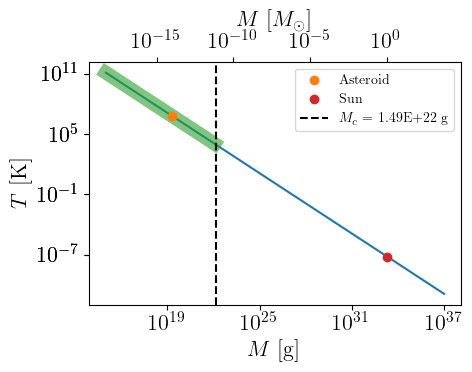}\label{fig:temperature_vs_mass}
\caption{{\bf(a)} 
PBHs evolution with redshift for specific temperatures of interest: Initial PBH masses for given blackbody temperatures in the redshift range of interest. In practice, the HR emitted from these BHs does not affect their mass evolution. 
{\bf (b)} Mass range with effective blackbody HR temperatures above the critical value ($T_{\rm crit} = 8000$): Relationship between Hawking temperature and BH mass. Data points show the Hawking temperature of some characteristic PBH masses. The green shaded region corresponds to effective blackbody temperatures above the critical temperature required to prevent the cloud from fragmenting. The dashed line quantifies the largest possible mass allowed by this constraint.}
\label{fig:mass_evolution_redshift}
\end{figure}

\subsection{PBH evolution with redshift}\label{sec:Ap_PBHs_evolution_redshift}

Following the discussion of Section \ref{sec:Hawking_radiation}, we show the evolution of BH masses due to HR in the redshift range of interest ($10\leq z \leq 30$) for four temperatures around the critical temperature ($T_{\rm crit}=8000$ K). 

Figure \ref{fig:mass_evolution_redshift} plots the initial mass of a PBH that would have a specific HR temperature as a function of redshift (found by solving eq. \ref{eq:T_evolution_HR}). For each temperature, the initial PBH mass needed is effectively the same for the entire range of redshifts. BH masses can only change significantly close to their evaporation time: for a PBH that would emit at around $T_{\rm crit}$ in the above redshift range, the evaporation time is estimated at $10^{24}$ Gyr, much larger than the age of the Universe.

Finally, we also note that for the mass ranges considered, PBH accretion is expected to be negligible \citep{Rice_Zhang_2017}.

\begin{figure}[h]
    \centering
    \includegraphics[width=0.68\linewidth]{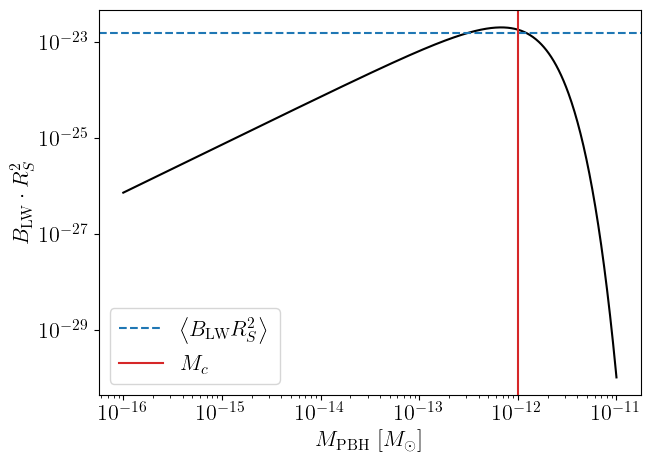}
    \caption{Change of $B_{\rm LW} R_S^2$ with PBH mass. The dashed line shows the average value from the same range of masses for a specific PBH mass function.}
    \label{fig:pbh_mass_function_effects}
\end{figure}

\subsection{PBH mass functions}\label{sec:Ap_mass_functions}

At the time of formation, PBHs can have a range of masses that may follow one of several proposed model-dependent mass distributions \citep{Chen_Hall_2024}. Here, we show that a population of monochromatic PBHs -- which is all that we consider in the main text -- is sufficient to test the viability of the scenario at hand.

To demonstrate this, we consider the factor $B_{\rm LW} R_S^2$ (the product of the specific intensity at the LW band with the squared Schwarzschild radius) since this factor plays the crucial role in determining specific intensity (see eq. \ref{eq:J_LW_single_BH}). It is interesting to note that these two terms have an inverse behavior as a function of mass: the specific intensity gets bigger as the mass is reduced, while the Schwarzschild radius grows linearly with mass. 

The mean value of this factor, $\left< B_{\rm LW} R_S^2 \right>$, over the relevant range of masses for some mass function {\it P(m)} is given by: 

\begin{equation}
    \left< B_{\rm LW} R_S^2 \right> = \int_{M_{\rm min}}^{M_{\rm max}} B_{\rm LW} R_S^2 \ P(m) dm.
\end{equation}

We can quantify the importance of a given mass function by comparing this mean value to the value of $B_{\rm LW} R_S^2$ calculated at each of the individual masses in the range. Figure \ref{fig:pbh_mass_function_effects} shows the $B_{\rm LW} R_S^2$ product over a range of masses compared to the mean value of this factor for an example mass function. For demonstrative purposes, the mass function considered here is a log-normal mass function\footnote{Where $P(m)$ is normalized to unity: $\int_0^\infty P(m) dm = 1$.}

\begin{equation}
    P_{\rm LN}(m) = \frac{1}{\sigma m \sqrt{2 \pi}} \exp \left( - \frac{\ln^2(m/M_c)}{2\sigma^2} \right),
\end{equation}

where $M_c$ denotes the median mass and $\sigma$ the width of the mass distribution. Here, we chose a median mass near the peak of the $B_{\rm LW} R_S^2$ product plotted in Figure \ref{fig:pbh_mass_function_effects}: $M_c = 10^{-12}\: \rm  M_\odot$. Note that this also explains the dip in Figure \ref{fig:Constant_D_monochromatic}(b). We also set $\sigma=0.5$ so that the distribution is not very broad; however, the choice of $\sigma$ does not affect the main conclusions.

Figure \ref{fig:pbh_mass_function_effects} demonstrates that a mass distribution, as expected from the properties of the mean, can outperform a number of individual masses but cannot surpass the maximum value of the $B_{\rm LW} R_S^2$ distribution. Hence, choosing a PBH mass distribution can act as an optimistic choice, but will not alter the constraints of the best monochromatic PBH scenario.

\bibliographystyle{mnras}
\bibliography{references.bib} 

\end{document}